%% file: main.tex
\documentclass[sigconf]{acmart}
\input{preamble.tex}
\AtBeginDocument{%
  \providecommand\BibTeX{{%
    Bib\TeX}}}

\setcopyright{none}
\renewcommand\footnotetextcopyrightpermission[1]{}
\begin{document}

\title{Aquas: Enhancing Domain Specialization through Holistic Hardware-Software Co-Optimization based on MLIR}

\author{Yuyang Zou}
\email{yyzou25@stu.pku.edu.cn}
\affiliation{%
  \institution{Peking University}
  \city{Beijing}
  \country{China}}

\author{Youwei Xiao}
\email{shallwe@pku.edu.cn}
\affiliation{%
  \institution{Peking University}
  \city{Beijing}
  \country{China}}

\author{Chenyun Yin}
\email{higgs@stu.pku.edu.cn}
\affiliation{%
  \institution{Peking University}
  \city{Beijing}
  \country{China}}

\author{Yansong Xu}
\email{yansongxu.mail@gmail.com}
\affiliation{%
  \institution{Peking University}
  \city{Beijing}
  \country{China}}

\author{Yuhao Luo}
\email{luoyuhao584@gmail.com}
\affiliation{%
  \institution{Peking University}
  \city{Beijing}
  \country{China}}

\author{Yitian Sun}
\email{ytsun@stu.pku.edu.cn}
\affiliation{%
  \institution{Peking University}
  \city{Beijing}
  \country{China}}

\author{Ruifan Xu}
\email{xuruifan@pku.edu.cn}
\affiliation{%
  \institution{Peking University}
  \city{Beijing}
  \country{China}}

\author{Renze Chen}
\email{crz@pku.edu.cn}
\affiliation{%
  \institution{Peking University}
  \city{Beijing}
  \country{China}}

\author{Yun Liang}
\email{ericlyun@pku.edu.cn}
\affiliation{%
  \institution{Peking University}
  \city{Beijing}
  \country{China}}

\begin{abstract}
Application-Specific Instruction-Set Processors (ASIPs) built on the RISC-V architecture offer specialization opportunities for various applications. Existing frameworks are largely designed around fixed instruction extension interfaces and rely on manual software adaptation. However, as emerging domains scale up in complexity, two major challenges arise.
First, memory access remains a primary bottleneck as existing design flows lack architectural awareness of memory interfaces, leading to suboptimal interface selection and orchestration. Second, the semantic complexity of custom instruction extensions, characterized by non-trivial control logic and irregular memory behaviors, hinders the ability of conventional compilers to perform automated and comprehensive offloading.

We present \apsmlir, a holistic hardware-software co-design framework built upon MLIR. \apsmlir proposes a memory interface model that jointly considers interface characteristics and cache effects, along with an interface-aware synthesis flow guided by this model that progressively optimizes the input specification and generates efficient hardware implementations.
We also propose an e-graph-based retargetable compiler approach with a novel matching engine for efficient instruction mapping and offloading, enabling robust and effective utilization of custom instruction capabilities.
Case studies across four diverse domains show that \apsmlir delivers substantial acceleration, achieving up to 15.61\x speedup with 14.5\% area overhead and zero frequency degradation, proving highly competitive in domain acceleration against more powerful general-purpose cores and vector extensions.
\end{abstract}

\maketitle

\input{sec/0_intro}
\input{sec/1_preliminary}
\input{sec/1.5_overview}
\input{sec/2_semantics_and_synthesis}
\input{sec/3_compiler}
\input{sec/5_casestudy}
\input{sec/9_conclusion}

\newpage

\bibliographystyle{ACM-Reference-Format}
\bibliography{refs}
\end{document}

%% file: preamble.tex
\usepackage[ruled,linesnumbered,vlined]{algorithm2e}
\usepackage{amsmath}
\usepackage[normalem]{ulem}
\usepackage{pifont}
\usepackage{titlesec}
\usepackage{listings}
\usepackage{subcaption}
\usepackage{float}
\usepackage{tikz}
\usepackage{multirow}
\usepackage{tabularray}
\usepackage{tabularx}
\usepackage{stackengine}
\usepackage[export]{adjustbox}
\usepackage{soul}
\usepackage{xspace}
\usepackage{graphicx}
\usepackage{booktabs}
\usepackage{etoolbox}
\usepackage{lipsum}
\usepackage{enumitem}
\usepackage{wrapfig}
\usepackage{hyperref}
\usepackage{makecell}
\usepackage{url}

\def\BibTeX{{\rm B\kern-.05em{\sc i\kern-.025em b}\kern-.08em
    T\kern-.1667em\lower.7ex\hbox{E}\kern-.125emX}}

\newlist{todolist}{itemize}{2}
\setlist[todolist]{label=$\square$}
\usepackage{pifont}
\newcommand{\GrayCircle}[1]{\raisebox{-0.45ex}{\text{\LARGE\color{darkgray}#1}}}
\newcommand{\NumberCircle}[1]{\GrayCircle{\ding{\the\numexpr 181+#1\relax}}}

\newcommand{\apsmlir}{\textit{Aquas}\xspace}
\newcommand{\egraph}{\textit{e-graph}\xspace}

\newcommand{\Egraph}{\textit{E-graph}\xspace}

\newcommand{\egglog}{\textit{egglog}\xspace}

\newcommand{\x}{\texttimes\xspace}
\newcommand{\compactparagraph}[1]{\emph{{\textbf{#1}}}}

%% file: sec/0_intro.tex
\section{Introduction}

Edge computing is increasingly driving the need for Application-Specific Instruction-Set Processors (ASIPs), which accelerate applications through hardware specialization using instruction set architecture extensions (ISAXs), while maintaining general software programmability. RISC-V~\cite{waterman_risc-v_2014}-based ASIPs have been widely adopted in signal processing~\cite{scheffler_saris_2024, gautschi_near-threshold_2017}, machine learning inference~\cite{armeniakos_mixed-precision_2025, yang_flexacc_2021, gerogiannis_deca_2025}, cryptography~\cite{pan_finesse_2025, li_scalable_2022, cheng_risc-v_2024}, and computer graphics~\cite{tine_skybox_2023, tine_vortex_2021}. Designing high-performance ASIPs presents a significant hardware-software co-design challenge, demanding a framework that can both generate specialized hardware and adapt software to leverage it effectively.

Current commercial tools on ASIP specialization, represented by Synopsys ASIP Designer~\cite{synopsys_inc_asip_2025}, allow designers to customize processors' internal microarchitecture at the cycle level for complete customization.
Research frameworks, such as Longnail~\cite{oppermann_longnail_2024} and APS~\cite{xiao_invited_2025}, extend existing RISC-V cores with instruction extension interfaces like RoCC~\cite{amid_chipyard_2020} and CV-X-IF~\cite{openhw_group_openhwgroupcore-v-xif_2026}, and synthesis execution units using high-level synthesis (HLS) tethered to these interfaces.
Users then write compiler rules by hand to offload software code to these specialized instructions.
These methodologies excel for basic ISAXs like scalar or packed SIMD.
However, emerging domains like Large Language Model (LLM) inference drive the demand for substantial computational power.
As the demand grows explosively, these frameworks all fall short under scaled design complexity.

Here, we summarize two indispensable challenges in current ASIP specialization: \textbf{Challenge 1: How to fully capture interface characteristics to improve the overall memory access efficiency of ISAXs}.
Emerging data-intensive applications impose immense demands on memory access bandwidth.
However, data movement in tightly-coupled ASIPs is inherently constrained by the core's memory hierarchy for coherency. This tight integration often becomes a bandwidth bottleneck, as hardware designers often blindly select and orchestrate memory access patterns without considering the interfaces' capabilities.
Commercial tools ~\cite{synopsys_inc_asip_2025} offer fine-grained customization, but their cycle-level nature conflates core hardware logic with specific interface timing. This lack of modularity results in a prohibitive engineering overhead.
Research frameworks ~\cite{oppermann_longnail_2024,xiao_invited_2025} are limited to rigid abstraction of their targeted instruction extension interfaces.
These frameworks also lack the architectural flexibility to integrate essential features like local scratchpads, which further worsens bandwidth limitations and confines them to basic ISAXs.
\textbf{Challenge 2: How to robustly discover ISAX offloading opportunities to fully exploit the acceleration potential provided by this hardware.}
The primary obstacle lies in the widening semantic gap between highly diverse software implementations and increasingly sophisticated ISAX semantics.
While application code exhibits significant syntactic divergence despite semantic equivalence, hardware specialization has also evolved to incorporate complex control flow and complex memory access patterns. This dual-sided complexity leads to an explosion in the program variant space.
Furthermore, current compiler technologies exacerbate this explosion through premature lowering of application code into low-level representations.
This process strips high-level semantics and introduces implementation-specific noise, which exponentially inflates the search space that the compiler must navigate.
Therefore, frameworks like ~\cite{xiao_invited_2025} can only rely on brittle matching rules that fail under minor code variations, especially loop transformations.
Even commercial tools~\cite{synopsys_inc_asip_2025} require manual specification between instruction behaviors and C intrinsic due to similar reasons.
In this work, we present \apsmlir, a holistic, MLIR-based~\cite{lattner_mlir_2021} framework for automated hardware-software co-design that generates optimized ASIPs and their corresponding compiler support.
On the hardware side, \apsmlir centers on a model of memory-interface attributes and cache effects, with Aquas-IR providing a multi-level intermediate representation to carry this model through the synthesis flow. The synthesis flow progressively selects access mechanisms and refines transaction granularity before lowering the result into implementation.
For the compiler, \apsmlir overcomes the vulnerability of traditional approaches to syntactic divergence by bridging the hardware-software semantic gap at a structural level. By representing the program within an \egraph~\cite{willsey_egg_2021} natively aligned with ISAX semantics, our approach leverages a synergistic combination of algebraic and control-flow rewrites to non-destructively accumulate a vast equivalence space of program variants. \apsmlir then introduces a hybrid skeleton-components matching engine that strategically navigates this equivalent space to identify mapping candidates robustly and efficiently, unearthing complex offloading opportunities that typically evade conventional compilers.
\apsmlir unifies the entire ASIP design flow within the MLIR infrastructure, enables seamless co-optimization across the hardware-software boundary, and facilitates agile design iteration.

Our contributions are as follows:
\begin{itemize}[leftmargin=*]
    \item We introduce \apsmlir, a holistic open-source framework for unified and automated ASIP hardware-software co-design built upon MLIR infrastructure.
    \item We propose an interface-aware synthesis flow that models microarchitectural memory interface attributes and automatically generates optimized hardware implementation.
    \item We design a novel retargetable compiler that couples MLIR’s structural representation with \egraph-driven equivalence exploration, facilitating robust mapping of application code to specialized ISAXs via a skeleton-components matching mechanism.
\end{itemize}

We showcase \apsmlir using four real-world case studies, including post-quantum cryptography, point-cloud processing, graphics rendering, and CPU LLM inference. Compared to the base core, \apsmlir achieves up to 15.61\x speedup on individual kernels, and 1.95\x on end-to-end workloads. The integration of these ISAXs incurs less than 22.9\% area overhead on end-to-end. Against BOOM~\cite{zhao_sonicboom_2020}, a high-performance out-of-order core, and Saturn~\cite{zhao_saturn_2024}, a RISC-V vector unit, it achieves competitive performance while saving 92.3\% and 34\% chip area, respectively, without frequency degradation, demonstrating its area efficiency for domain-specific workloads.

%% file: sec/1_preliminary.tex
\section{Preliminaries}
\label{sec:preliminaries}

\subsection{Instruction Extension Interfaces}

Instruction extension interfaces, including RoCC ~\cite{amid_chipyard_2020}, CV-X-IF~\cite{openhw_group_openhwgroupcore-v-xif_2026}, and SCAIE-V~\cite{damian_scaie-v_2022}, provide a convenient way to implement custom instructions without modifying the base processor microarchitecture.
They expose the fundamental interfaces needed for ISAX specialization, including instruction offloading, register-file access, and memory access.
Prior frameworks like \cite{xiao_invited_2025, oppermann_longnail_2024} rely entirely on these interfaces and directly generate hardware targeting them.

As emerging workloads such as LLM inference increase memory pressure, interface limitations become a primary bottleneck.
These interfaces' memory ports are constrained by limited width and inevitable bubbles, resulting in a low effective bandwidth.
Designers, therefore, need to route some transfers through other high-bandwidth interfaces, such as the system bus, depending on the access pattern and performance target.
Prior ASIP flows usually handle this tradeoff with manual, first-glance choices.
However, \textit{suboptimal memory interface selection and deficient orchestration can lead to marked performance loss}, motivating us to explore a systematic approach to interface-aware memory access optimization.

\subsection{MLIR Infrastructure}

MLIR~\cite{lattner_mlir_2021} provides a powerful solution for optimizing this hardware interface problem. It is a modular compiler infrastructure that allows defining intermediate representations at various levels of abstraction and constructing analyses and transformations for optimizations at the appropriate levels. MLIR can also further handle hardware synthesis tasks. CIRCT~\cite{eldridge_mlir_2021}, built on top of MLIR, provides concrete support for this. Moreover, HLS frameworks HECTOR ~\cite{xu_hector_2022}, ScaleHLS~\cite{ye_scalehls_2022}, and Dynamatic~\cite{josipovic_dynamatic_2025}, and ASIP synthesis framework Longnail ~\cite{oppermann_longnail_2024}, are also built with MLIR.

MLIR's rich expressiveness allows ISAX to encompass various access patterns and control flows beyond simple dataflow. However, this increased design complexity also poses a new challenge for software compilers: \textit{Compilers need to robustly analyze these complex access patterns and control flows to discover as many ISAX offloading opportunities as possible to fully leverage the performance gains}. Fortunately, MLIR is also a strong infrastructure for compilers, with layers of abstraction that allow us to handle both hardware and software at the same level, helping to solve this challenge.

\subsection {\Egraph Representation}

An \egraph~\cite{zhang_better_2023} is a data structure that compactly represents a large number of equivalence relations, and serves as a powerful tool to assist MLIR in solving this software matching challenge. Within an \egraph, \textit{e-classes} group semantically equivalent \textit{e-nodes} together. An \textit{e-node} is a function symbol applied to child identifiers, where each identifier denotes a child \textit{e-class}. Rewrites can be performed on the \egraph, which match a pattern of \textit{e-nodes} and merge the rewrite results into the original \textit{e-class} through \textit{union} operations. By continuously rewriting, the \egraph can maintain an equivalent space that preserves all rewrite results. An extraction step selects the \textit{e-nodes} that minimize a user-defined cost function, generating a set of optimal expressions.

\medskip

Overall, we recognize two fundamental challenges in current ASIP specialization:
\textit{systematically explore interface-aware synthesis decisions}, and
\textit{robustly discover ISAX offloading opportunities}.
Our insight is to provide a unified toolchain for modeling and exploring hardware synthesis, while solving the software robust retargeting problem with \egraph integration. We use various MLIR infrastructures and assets to achieve complete flow integration.

%% file: sec/1.5_overview.tex
\section{Unified Toolchain}
\label{sec:overview}

\input{fig_tex/tool_fig.tex}

As illustrated in \autoref{fig:toolchain}, \apsmlir contributes a unified toolchain for agile, flexible, and fully optimized ASIP design. The entire flow deeply integrates MLIR's dialects and assets, and introduces new abstractions to build a cross-level synthesis and compilation pipeline.
For the hardware-side challenge, we propose a model that jointly considers interface characteristics and cache effects (\autoref{sec:model}), and an intermediate representation (\textit{Aquas-IR}) to symbolize and guide decisions during synthesis (\autoref{sec:ir}). We propose an interface-aware synthesis flow that progressively optimizes the input specification and generates efficient hardware implementation (\autoref{sec:synthesis}).
For the software-side challenge, we represent the program in MLIR with a similar abstraction level to the ISAX functional behavior (\autoref{sec:compiler-cano}), and use \egraph to expand its equivalence space (\autoref{sec:compiler-egraph}). By hybrid applying internal and external rewrites under ISAX structure's guidance, we fully expose ISAX offloading opportunities (\autoref{sec:compiler-rewrite}). Targeting these ISAXs, we apply the skeleton-component matching technique to robustly and accurately locate potential offload points, enabling fully automated retargetable compilation (\autoref{sec:compiler-matching}). With this flow, Aquas enables holistic hardware-software co-optimization for domain specialization.

%% file: fig_tex/tool_fig.tex
\begin{figure}[t]
  \centering
  \includegraphics[width=0.8\linewidth]{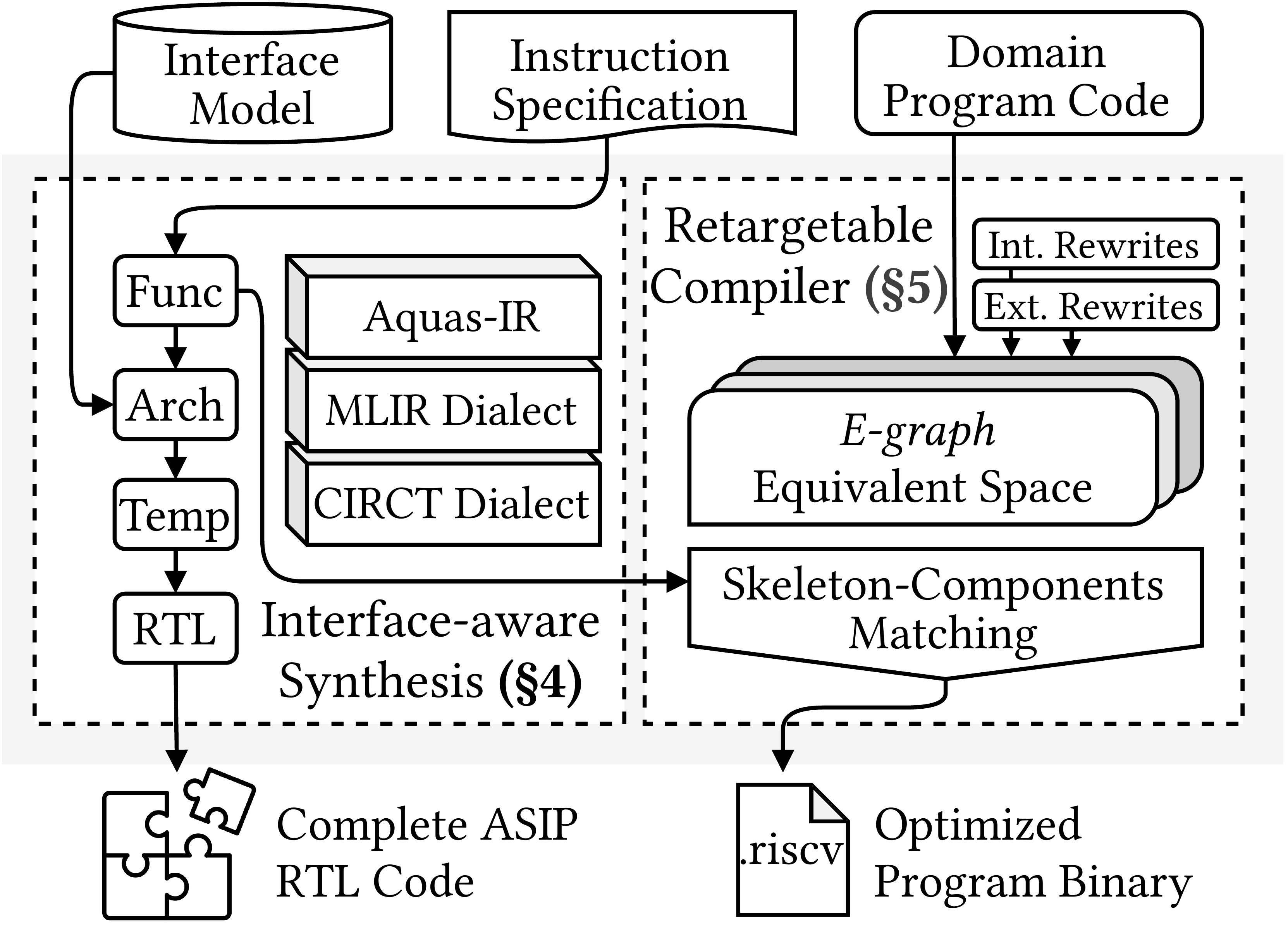}
  \caption{
    Overview of the unified toolchain in \apsmlir.
  }
  \label{fig:toolchain}
\end{figure}

%% file: sec/2_semantics_and_synthesis.tex
\section{Modeling and Synthesis}
\label{sec:model_and_synthesis}

\input{fig_tex/interface-limitation}

This section presents \apsmlir's interface-aware hardware synthesis, which is structured around a core-ISAX interface model, a progressive multi-level representation (\textit{Aquas-IR}), and a synthesis-time optimization pipeline to generate the final hardware design.

\subsection{Modeling Core-ISAX Interfaces}
\label{sec:model}

The choice of memory access mechanism can have a significant impact on overall performance. Consider the interfaces depicted in \autoref{fig:interface-limitation}(a). The \texttt{@itfc1}, typically provisioned by an instruction extension interface, provides low access latency but is constrained by a narrow 32-bit width, lack of burst support, and a limit of a single in-flight transaction. Conversely, \texttt{@itfc2} offers a wide 64-bit width, burst transfer capabilities, and supports up to two in-flight transactions, but with higher latency. Without an analytical model, hardware designers are prone to making suboptimal design decisions. As shown in \autoref{fig:interface-limitation}(b), even minor design decisions, such as improper interface selection or access ordering, can introduce a notable 7- to 9-cycle latency penalty, and may severely degrade overall performance in larger designs.

To systematically analyze these performance implications, we propose a novel core-ISAX interface model. In this model, each memory interface $k$ can be expressed as a 6-tuple:
\[
(W_k,\; M_k,\; I_k,\; L_k,\; E_k,\; C_k),
\]
where $W_k$ is the interface width in bytes, $M_k$ is the maximum beat count of one transaction, $I_k$ is the maximum in-flight transactions, $L_k$ and $E_k$ denote read lead-off latency and write completion cost, and $C_k$ is the cache-line size visible to that interface.

These interfaces must adhere to specific \textbf{microarchitectural constraints}.
A transaction of size $m$ is legal if its beat count $m/W_k = 2^t \le M_k$ for some $t \in \mathbb{N}_0$. Additionally, the starting address must be aligned to $m$, and both read and write transactions support independent pipelining up to the in-flight transaction limit $I_k$.

Based on this model, the total latency for a sequence of $N$ load or $N$ store transactions on interface $k$ can be estimated as follows. Let $m_j$ be the size in bytes of the $j$-th transaction. We define $a_j$ as the issue cycle and $b_j$ as the completion cycle of the transaction, with $a_j = b_j = -1$ for $j \le 0$. $a_j$ and $b_j$ are computed as:
\[
\begin{aligned}
a_j &= 1 + \max(a_{j-1}, b_{j-I_k}), \\
b_j^{ld} &= \frac{m_j}{W_k} + \max(b_{j-1}, a_j + L_k - 1), \\
b_j^{st} &= \frac{m_j}{W_k} + E_k + \max(b_{j-1}, a_j - 1).
\end{aligned}
\]

These recurrences serialize transactions waiting for structural slots while overlapping data transfers. The completion cycle $b_N$ gives the estimated interface latency for the sequence.

\compactparagraph{Cache Hierarchy and Locality.}
Cache effects are also non-negligible during synthesis. \apsmlir accounts for these through two primary mechanisms. First, we incorporate the cache-line size $C_k$ into the model to mitigate cache thrashing during execution. Second, we introduce \texttt{cache\_hint} label for memory interfaces and scratchpad buffers declared in ISAX to prevent hierarchy mismatches that incur costly synchronization cycles. These hints are often readily inferable. Take a finite impulse response (FIR) filter as an example, the large coefficient vector can be tagged as \texttt{cold} as these are typically read directly from DRAM, while CPU-initialized configuration parameters are tagged as \texttt{warm} to favor higher-level paths.

\subsection{Aquas-IR}
\label{sec:ir}

\input{table/aquas_ir}

To seamlessly integrate this modeling space into the synthesis workflow, we introduce Aquas-IR. Aquas-IR is organized as three refinement levels, where each level captures a specific aspect of the synthesis process, as summarized in \autoref{tab:aquasir}.

At the \textbf{functional level}, operations are close to high-level software abstractions. For example, \texttt{transfer} and \texttt{fetch} are agnostic to the underlying access mechanism, and simply specify the source, destination, and size of the transfer.
The \textbf{architectural level} makes decisions regarding the physical access mechanism for memory operations that require explicit interface binding.
We introduce \texttt{!memitfc<>} as module-level symbols that encapsulate the physical interface model defined in \autoref{sec:model}. Each \texttt{memitfc} symbol is associated with a unique name and a set of interface parameters $W, M, I, L, E$ and $C$. This information will be extracted during synthesis to guide interface binding and transfer length decomposition. Operations at this level, such as \texttt{copy} and \texttt{load}, are explicitly bounded to one physical interface, and annotated with the corresponding \texttt{memitfc} symbol. This also means that the legality of these operations is now subject to the microarchitectural constraints of the chosen interface. Finally, the \textbf{temporal level} defines the order of these decomposed transactions, accounting for the in-flight limit $I$ and the cache-line size $C$. Additionally, to account for possibly multi-cycle memory latency, we model transactions as asynchronous issue and wait pairs. The order of these operations is guaranteed by the \texttt{after} attribute of each operation.

Aquas-IR is implemented as a specialized MLIR dialect, leveraging its robust infrastructure for arithmetic, control flow, and affine transformations to serve as a versatile intermediate target for instruction description languages like CADL\cite{xiao_invited_2025}.

\subsection{Synthesis-Time Optimization}
\label{sec:synthesis}

\input{fig_tex/sequence.tex}

\input{fig_tex/synthesis.tex}

Guided by the analytical model, \apsmlir performs an interface-aware synthesis optimization that progressively optimizes and lowers design specifications through Aquas-IR. Using the \texttt{fir7} example, where naive manual designs often result in suboptimal schedules like \autoref{fig:sequence}(a), we demonstrate how this optimization identifies a faster schedule (b), with corresponding intermediate representation (IR) refinements illustrated in \autoref{fig:synthesis}.

\compactparagraph{Scratchpad Buffer Elision.}
ISAXs often explicitly stage data in local scratchpads. \apsmlir first evaluates whether these intermediate buffers can be safely elided to allow direct main memory access, possibly reducing latency and SRAM usage. This optimization is applied on the functional level IR. To prevent extra latency overhead or cache degradation, we disable elision for scratchpads accessed within unrolled regions, outside pipelined loops, or used purely as local temporary buffers. For the remaining candidates, affine analysis is applied to avoid elisions that trigger cache thrashing. The transformation is ultimately accepted only if tentative loop rescheduling confirms no overall latency increase.
For our example in \autoref{fig:sequence}, we elide the scratchpad \texttt{bias} according to previous analysis. After this optimization, the bulk transfer latency is eliminated, and the newly introduced per-element access latency of each \texttt{bias[i]} can be effectively hidden by the computation of accumulation, making this elision beneficial.
In the IR, the declaration for \texttt{bias} is removed, and the original \texttt{read\_smem} operation is replaced with global memory \texttt{fetch}, shown in \autoref{fig:synthesis}(a).

\compactparagraph{Interface Selection and Canonicalization.}
This step lowers functional-level memory operations to the architectural level by solving an optimization problem that assigns each memory operation to exactly one visible memory interface.
Consider all reads or all writes separately within one region. For each operation $q$ of $m_q$ bytes, \apsmlir chooses exactly one interface $k$ to implement this transaction, denoted by the binary variable $X(q,k)$, and greedily splits the request into legal transfer sizes of interface $k$ in decreasing order. For a properly aligned base operation, this decomposition yields an ordered sequence of naturally aligned, legal transfers $\{m_{q,p}^{(k)}\}_p$. The optimization can be formulated as:
\[
\min \sum_k T_k + \sum_{q,k} X(q,k)\left\lceil \frac{m_q}{C_k} \right\rceil \cdot \frac{C_k}{W_k}.
\]
where $T_k$ is the estimated transfer latency on interface $k$ based on the model, and the second term penalizes cache hierarchy mismatch by approximating the beat count needed to synchronize the touched cache lines. Here, for each interface $k$ that has operations assigned, $T_k$ can be estimated using an approximation model as follows:
\[
T_k =
\begin{cases}
L_k - 1 + \sum_q X(q,k)\left[\sum_p \max\!\left(\frac{L_k}{I_k}, \frac{m_{q,p}^{(k)}}{W_k}\right)\right], & \text{if Ops are load}\\
\sum_q X(q,k)\left[\sum_p \left(\frac{m_{q,p}^{(k)}}{W_k} + E_k\right) \right]- 1, & \text{if Ops are store}
\end{cases},
\]
where $\frac{L_k}{I_k}$ simulates the bubbles introduced due to limited $I_k$.

In \autoref{fig:sequence}, the naive manual design suboptimally assigns the large \texttt{src} buffer to the narrow \texttt{@cpuitfc} interface. By evaluating latency and cache penalties, \apsmlir instead routes \texttt{src} through the high-bandwidth \texttt{@busitfc} for a lower cost, which eventually succeeds in reducing the latency of it. At the IR level, shown in \autoref{fig:synthesis}(b), \apsmlir lowers \texttt{transfer} operations into interface-bound \texttt{copy} operations, where \texttt{copy} specifies the corresponding interface. For \texttt{src}, to satisfy \texttt{@busitfc}'s constraints, the 108-byte transaction is canonicalized into 64-, 32-, 8-, and 4-byte legal transfers, symbolized by four \texttt{copy} operations.

\compactparagraph{Transaction Scheduling and Ordering.} This step lowers the architectural-level transfers to the temporal level by selecting the transaction order that minimizes completion time under finite $I_k$ and hierarchy constraints. \apsmlir first group these transfers using cache hierarchy levels. For read operations, transfers closer to the top of the hierarchy are issued earlier to prevent cold data from evicting hot data in the cache. For write operations, transfers closer to the bottom of the hierarchy are issued earlier to ensure hot data remains in the cache longer. Moreover, decomposed segments originating from the same memory operation are constrained to remain contiguous. Bounded by these rules, \apsmlir efficiently finds the minimal-latency schedule via a memoized search that compresses the exploration state into a relative timing window, exploiting the fact that latency recurrences are insensitive to global time translation. \apsmlir then lowers these ordered segments into asynchronous primitives, where each transfer becomes a \texttt{copy\_start}, and explicit \texttt{after} dependences capture the chosen issue order.

\autoref{fig:synthesis}(c) demonstrates the corresponding IR transformation. The four decomposed \texttt{copy} operations for \texttt{src} are optimally reordered and lowered into asynchronous \texttt{copy\_start} primitives, using \texttt{after} attributes to strictly encode their execution sequence. A final \texttt{copy\_wait} is then inserted to synchronize the last transfer, guaranteeing safe data access for subsequent computations.

\compactparagraph{Hardware Generation.}
After scheduling is fixed, \apsmlir transform each ISAX into a dynamic pipeline architecture following transactional semantics~\cite{hoe_synthesis_2000}.
Resource conflicts are automatically resolved at this step by inserting arbitration logic across each pipeline stage.
It then generates backend adapters for the instruction extension interface, as well as memory-access engines for handling protocol conversion, burst transfers, and runtime fallback handling for misaligned requests.
For explicit scratchpad buffers, \apsmlir also synthesizes multi-banked memory and its corresponding address mapping logic.
Finally, the resulting design is lowered to FIRRTL and SystemVerilog through CIRCT~\cite{eldridge_mlir_2021}.

%% file: fig_tex/interface-limitation.tex
\begin{figure}[t]
  \centering
  \includegraphics[width=\linewidth]{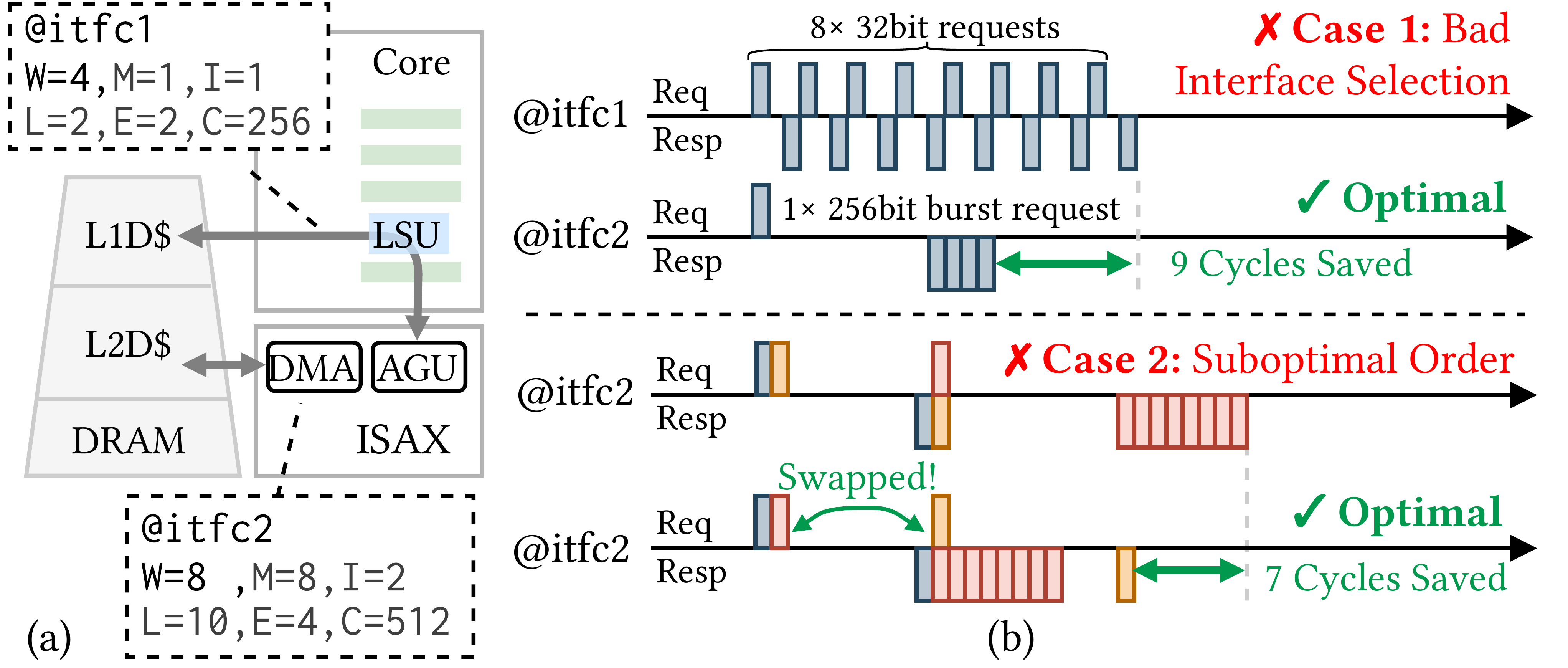}
  \caption{
    An example of available ISAX memory interfaces in an ASIP and common suboptimal design choices.
  }
  \label{fig:interface-limitation}
\end{figure}

%% file: table/aquas_ir.tex
\begin{table}[t]
\caption{Representative operations and exposed microarchitectural features across Aquas-IR abstraction levels.}
\label{tab:aquasir}
\centering
\small
\begin{tabularx}{\columnwidth}{p{0.13\columnwidth} p{0.29\columnwidth} p{0.48\columnwidth}}
\hline
\textbf{IR Level} & \textbf{Representative Ops} & \textbf{$\mu$-arch Features} \\ \hline
\multirow{3}{*}{Functional}
& \texttt{transfer, fetch}
& \multirow{3}{=}{$m$: Specify transfer size} \\
\cline{2-2}
& \texttt{read\_smem}
& \\
\cline{2-2}
& \texttt{read\_irf}
& \\ \hline
\multirow{3}{*}{\shortstack[l]{Architec-\\tural}}
& \texttt{!memitfc<...>}
& \multirow{3}{=}{$W, M_{\max}$: Determine legal length \newline $I, L, E$: Estimate load/store latency \newline $C$: Penalize cache refill/flush latency} \\
\cline{2-2}
& \texttt{copy\xspace\color{gray}{\# bulk}}
& \\
\cline{2-2}
& \texttt{load\xspace\color{gray}{\# scalar}}
& \\ \hline
\multirow{3}{*}{Temporal}
& \texttt{copy\_issue/wait}
& \multirow{3}{=}{$I$: Determine inflight-aware order \newline Cache hierarchy level: Determine phase order and cache locality} \\
\cline{2-2}
& \texttt{load\_issue/wait}
& \\
\cline{2-2}
& \texttt{read\_smem\_issue}, \texttt{read\_smem\_wait}
& \\ \hline
\end{tabularx}
\end{table}

%% file: fig_tex/sequence.tex
\begin{figure}[t]
  \centering
  \includegraphics[width=\linewidth]{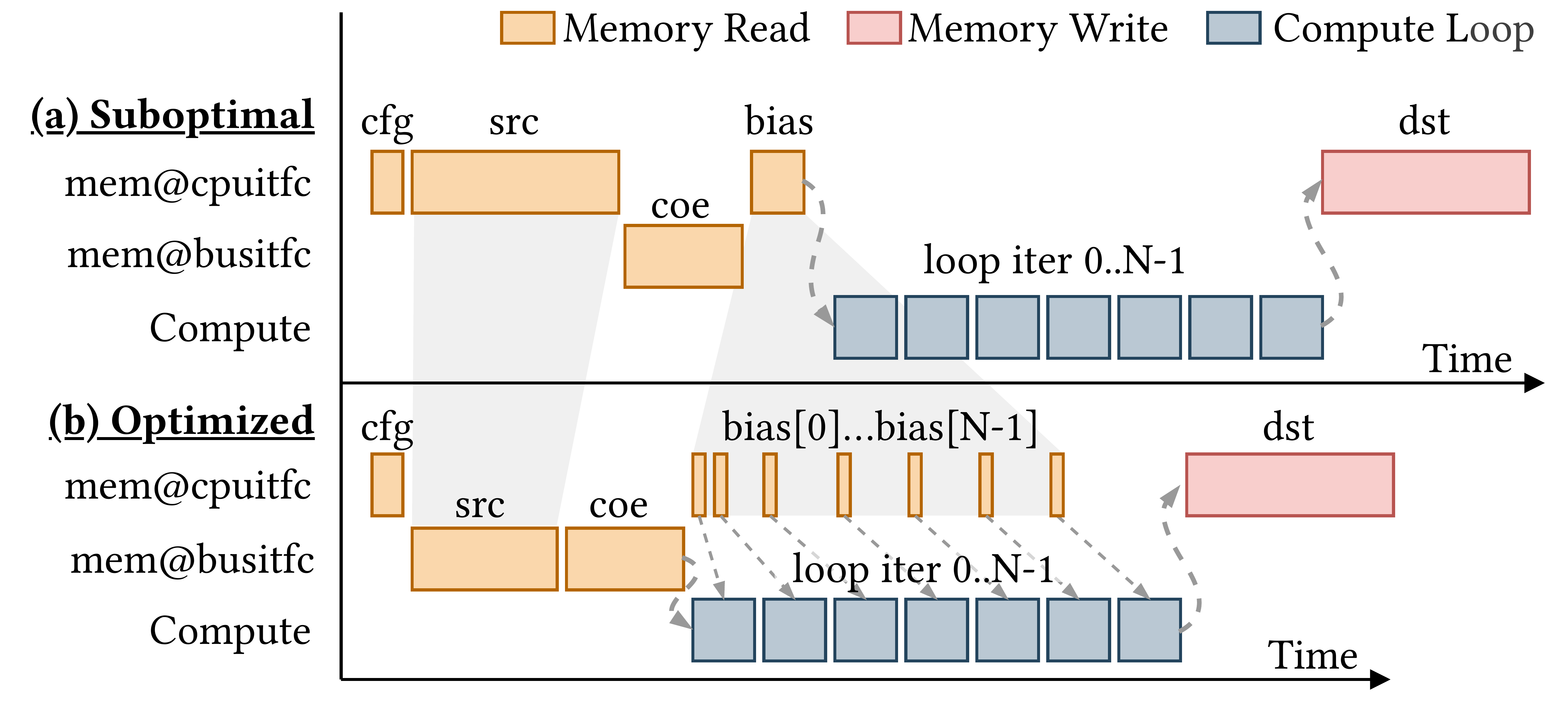}
  \caption{
    Timing diagram of the \texttt{fir7} kernel under a suboptimal lowering and under optimized synthesis pipeline.
  }
  \label{fig:sequence}
\end{figure}

%% file: fig_tex/synthesis.tex
\begin{figure}[t]
  \centering
  \includegraphics[width=\linewidth]{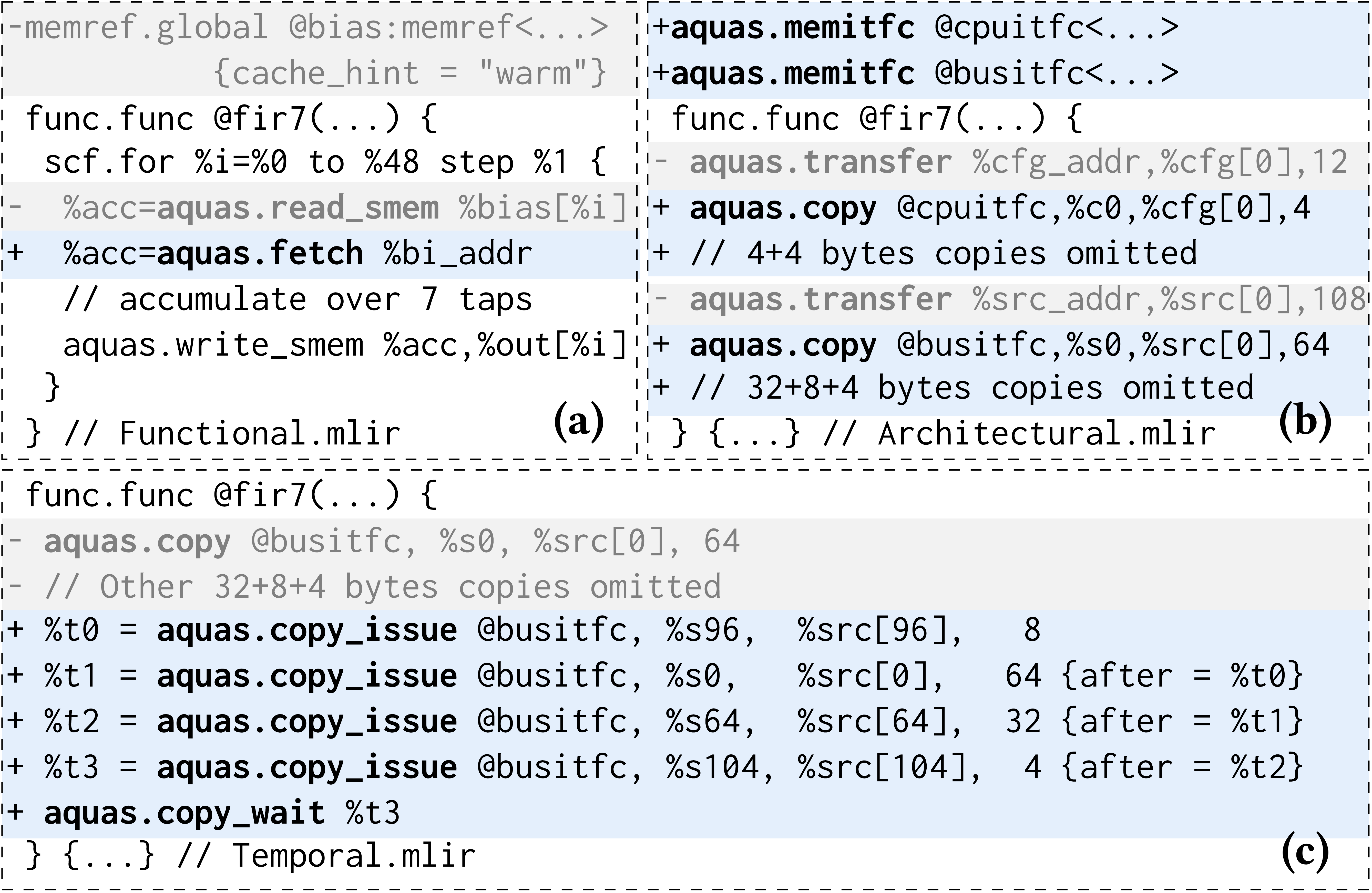}
  \caption{
    A synthesis example for the \texttt{fir7} kernel. (a) Scratchpad buffer elision. (b) Interface selection and canonicalization. (c) Transaction scheduling and ordering.
  }
  \label{fig:synthesis}
\end{figure}

%% file: sec/3_compiler.tex
\section{Retargetable Compiler}

\label{sec:compiler}

\input{fig_tex/compiler_e2e.tex}

This section presents \apsmlir's retargetable ASIP compilation flow. As shown in \autoref{fig:compiler_e2e}, it aligns hardware and software at a common abstraction level, performs hybrid rewriting on the \egraph to expand the software's equivalence space, and applies skeleton-component pattern matching for robust ISAX offloading.

\subsection{Semantic Alignment}
\label{sec:compiler-cano}

The fundamental obstacle to ISAX matching is the abstraction mismatch between software and hardware representations. Software semantics are expressed in terms of values, control flow, and memory effects, whereas ISAX hardware descriptions contain more microarchitecture-specific details, such as scratchpad access and register-file interactions. To enable reliable matching, we canonicalize both sides to a common abstraction level in base MLIR dialects.

As depicted in \autoref{fig:compiler_e2e}(\NumberCircle{1}), software code is translated into the base MLIR dialects through Polygeist, while hardware instruction descriptions are normalized from the functional level of Aquas-IR to the same base MLIR abstraction. Specifically, \textbf{Register operations} like \texttt{read\_irf} are eliminated by converting explicit register references into direct data dependencies. \textbf{Memory operations} like \texttt{transfer} are replaced with direct CPU memory accesses, and scratchpad buffers \texttt{A} and \texttt{C} are removed.
The translated ISAX form, therefore, retains only software-visible control flow and memory effects, making it structurally comparable to application code.

\subsection{Fusing MLIR and \Egraph Representations}
\label{sec:compiler-egraph}

The diversity of possible syntactic divergence makes a single MLIR program variant insufficient to cover all matching opportunities. We therefore introduce \egraph to the compiler flow as a structure that compactly represents equivalent program variants. In this flow, MLIR remains the structural backbone of the intermediate representation, while the \egraph enables large-scale exploration of equivalent variants in a non-destructive manner. We establish an integration scheme that maps MLIR structure and transformations into the \egraph while strictly preserving \textit{critical semantic relations} such as memory effects and control dependencies, which are often overlooked by other \egraph optimization solutions.

\compactparagraph{Encoding MLIR Programs into \Egraph. } We map each MLIR operation to an \textit{e-node} whose children correspond to the \textit{e-classes} of its operands. To encode strict execution ordering to \egraph, we separate the operations of each MLIR block into \emph{anchors} and dataflow operations. Anchors are operations that impose strict ordering constraints, including terminators (e.g., \texttt{yield}), side-effecting operations (e.g., \texttt{store}), and structured control flow. We structurally encode an entire MLIR block as a \texttt{tuple}(\ldots) e-node, using its anchors as direct children arranged in exact program order. The remaining operations form dataflow subtrees attached beneath the specific anchors that consume their results. In this way, the tuple captures the block-level sequencing skeleton, while the subtrees encode order-independent computations. This encoding natively preserves MLIR ordering and dominance within the \egraph.

\autoref{fig:compiler_e2e}(\NumberCircle{2}) shows the \egraph construction process. The white e-nodes represent the initially created \egraph. For brevity, we merge the tuple e-node for the loop region into its corresponding operation, denoted by bold arrows. The child e-nodes of the \texttt{for} loop, from left to right, represent the iteration step, the end bound, and the anchors (we omit the start bound for brevity). In this example, the outermost \texttt{for} loop in the software \egraph contains two anchors: the nested \texttt{for} loop and a \texttt{store} e-node.

\compactparagraph{Reuse MLIR Passes in \Egraph. }
\apsmlir enables MLIR's rich transformation infrastructure to be reused in tandem with the \egraph.
Whenever an MLIR pass needs to be invoked, the target program subgraph will be extracted first.
This step requires a cost model to select one optimal e-node from each e-class.
We then convert this extracted subgraph into MLIR operations to compose a valid MLIR program, which is subsequently fed into the selected MLIR passes. Upon completion, the modified MLIR is translated back into the \egraph as new e-nodes, and we perform a union operation to merge them with their original e-classes. The net effect is that the results of the MLIR transformations accumulate within the \egraph structure, rather than destructively overwriting previous states as in a traditional MLIR pass pipeline.

\subsection{Hybrid Rewriting for Space Expansion}
\label{sec:compiler-rewrite}

With \egraph representation, we can apply rewrites on software code to expand the search space of equivalent programs. We propose a hybrid rewriting strategy that iteratively applies two types of rewrites to the same \egraph until saturated.

The first type is \textbf{internal rewrites}, which apply dataflow transformations such as algebraic simplification. These rewrites focus on optimizing the dataflow subtrees beneath anchor e-nodes, while keeping the program's control flow structure intact.
As these rewrites do not touch the anchor nodes, the correctness of the program's control flow and side effects is preserved.
These rewrites are pre-defined as fixed \egglog~\cite{zhang_better_2023} rewrite rules.

The second type is \textbf{external rewrites}, which are control-flow-oriented and perform restructuring of the program's control flow, such as loop transformations including tiling and unrolling. These rewrites are difficult to hard-code as fixed rules. Moreover, the control flow they target often requires complex condition checks involving dependencies, ordering, and dominance information. Therefore, we implement these rewrites through reusing MLIR passes with the methodology in ~\autoref{sec:compiler-egraph}.

Blindly saturating external rewrites would cause the \egraph to grow explosively, as these rewrites often significantly reshape program structure. Such structural divergence can limit e-node sharing and introduce a multitude of new e-classes. To mitigate this, we employ an \textbf{ISAX-guided rewrite strategy} that prunes the search space by selectively triggering rewrites by analyzing target loop characteristics, effectively suppressing \egraph blowup. We explain this hybrid approach with \autoref{fig:compiler_e2e}(\NumberCircle{3}). The original software \egraph contains a nested loop with a non-affine index (\texttt{i$\ll$2}). A prior internal rewrite has already applied \texttt{i$\ll$2}$\rightsquigarrow $\texttt{i*4} rewrite on this. In the external rewrite stage, we first analyze the loop characteristics of the target ISAX, including iteration counts and stepping conditions. Based on these characteristics, we enumerate loop structures in the software code and compare their control flow differences. If they can be transformed through certain loop transformations, we apply them with MLIR passes through the method in ~\autoref{sec:compiler-egraph}.
The decision here only depends on the loop structure, not the specific operations within the loop body.
The cost model used here is defined as a heuristic scoring function that penalizes non-affine operations. By assigning a lower cost to affine-friendly expressions (e.g., choosing \texttt{i*4} over \texttt{i$\ll$2}), the extracted \egraph can be oriented toward more aggressive loop optimization. In this example, the inner loop is applied with unrolling with a factor of 2 after a series of analyses. The newly added e-nodes and edges are marked in red. Through this process, the final program structure exposes a direct equivalence with the target ISAX.

\subsection{Skeleton-Components Pattern Matching}
\label{sec:compiler-matching}

At the final step, we decompose each ISAX into a \textit{skeleton} and a few \textit{components} to enable reliable and scalable matching within this expanded search space. The skeleton captures the program's control structure and ordering constraints and is used to enforce correct execution order and loop-carried dependencies among key operations in each block. Components serve as dataflow subtrees beneath each anchor e-node.

During matching, we first generate \egglog tagging rules for each component. When a component's subtree is successfully matched in the \egraph, the rule rewrites it by inserting a unique marker e-node into the corresponding e-class. In the next step, a skeleton matching engine then identifies candidate e-classes whose enclosing blocks satisfy the required loop/region structure and contain the complete component set, guided by the decomposed instruction skeleton. It then performs a series of checks, including ordering, dominance/visibility, loop-carried dependencies, and effect constraints.
On success, another \texttt{ISAX} marker e-node is inserted into the matched e-class. As shown in \autoref{fig:compiler_e2e}(\NumberCircle{4}), the ISAX (bottom-right) is disassembled into a two-level loop skeleton and two \texttt{components}.
The compiler identifies two e-classes in the software \egraph that match the components above, and tags them with \texttt{Comp1} and \texttt{Comp2}. The skeleton matching engine then selects the outermost \texttt{for} node in the software \egraph, whose loop structure satisfies the skeleton due to previous rewrites. After validating the loop order and dependences, a \texttt{ISAX} e-node is inserted into the matched e-class.

After all these processes, \apsmlir extracts a single optimized program from the \egraph (\autoref{fig:compiler_e2e}(\NumberCircle{5})) using a cost model that prioritizes more \texttt{ISAX} e-nodes. \texttt{ISAX} markers then get transformed into intrinsic calls to the corresponding custom instructions. Finally, this lowered representation is fed into the standard MLIR-LLVM backend to generate the executable. This completes the end-to-end compilation flow, delivering a fully accelerated binary that seamlessly harnesses the targeted ASIP extensions.

%% file: fig_tex/compiler_e2e.tex
\begin{figure*}[t]
  \centering
  \includegraphics[width=0.85\linewidth]{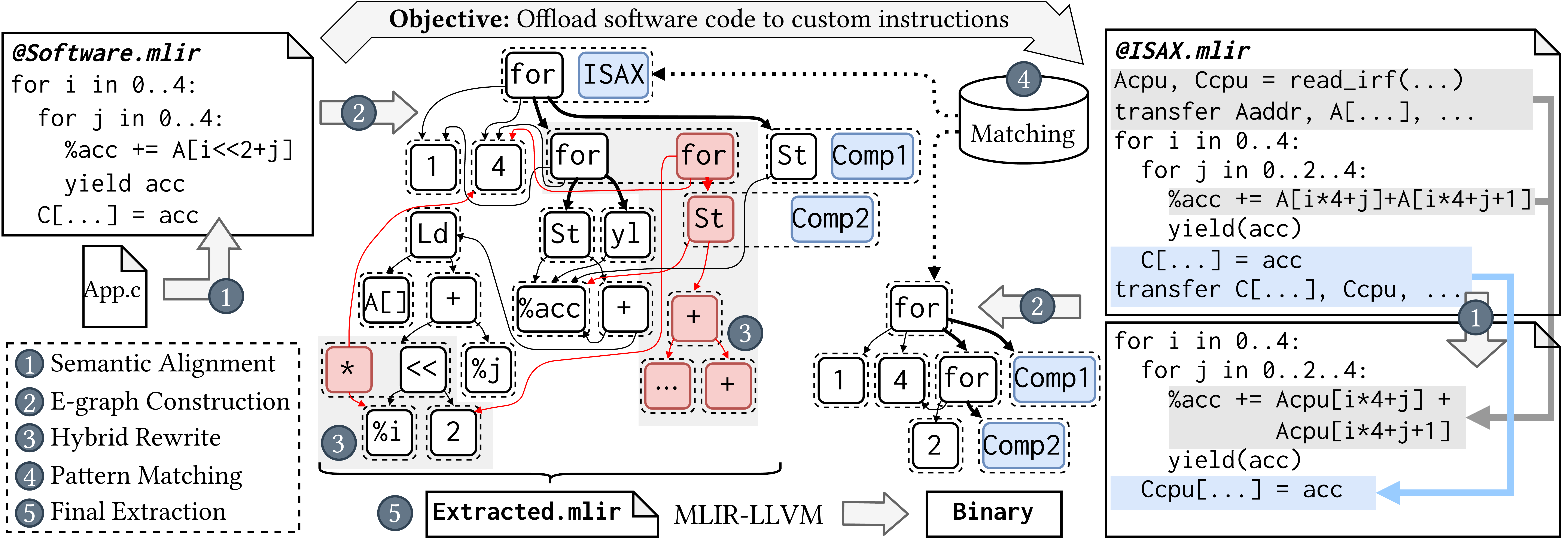}
  \caption{
    The end-to-end workflow of the \apsmlir's retargetable compiler.
  }
  \label{fig:compiler_e2e}
\end{figure*}

%% file: sec/5_casestudy.tex
\section{Case Studies}
\input{table/eval_pcl_deca}

We conduct case studies across four domains to evaluate \apsmlir in accelerating real-world workloads and its practical viability for physical silicon deployment.

\subsection{Experiment Setup}

We use Rocket core\cite{asanovic_rocket_2016} as our base processor core, on which \apsmlir is provided as a Chipyard~\cite{amid_chipyard_2020} plugin to generate the full ASIP RTL.
We enabled two memory access paths, one over the RoCC interface and the other over the system bus.
Our competing baselines include: (1) the base Rocket core integrated with ISAXs generated by the APS~\cite{xiao_invited_2025}, (2) the same base core augmented with Saturn~\cite{zhao_saturn_2024}, a mature general-purpose vector unit, and (3) the high-performance out-of-order BOOMv3 core~\cite{zhao_sonicboom_2020} without ISAXs. Area and timing metrics for ASIC are obtained using a commercial tool targeting a 130nm process at the \textit{RocketTile} level, and the baseline Rocket can achieve a maximum of 232MHz with an area of 4.11mm$^2$. We collect performance metrics via RTL simulation using Verilator~\cite{veripool_veripool_2025} and calculate speedups with maximum achievable frequency.

\subsection{Post-Quantum Cryptography}
\label{sec:pqc}

Syndrome computation is a core operation in code-based PQC, calculating $\mathbf{s}=\mathbf{He^\top}$ over $\mathbb{GF}(2^n)$ to verify or decrypt data using a parity-check matrix $\mathbf{H}$ and error vector $\mathbf{e}$. In practice, $\mathbf{H}$ is typically sparse, and $\mathbf{e}$ from multiple requests can be packed into a matrix to utilize matrix multiplication over $\mathbb{GF}(2^n)$. Accordingly, we design \texttt{vdecomp} for bistream unpacking and \texttt{mgf2mm} for computation.

\autoref{tab:eval_pcl_deca} demonstrates that \apsmlir achieves substantial speedups of 7.59\x for \texttt{vdecomp} and 3.29\x for \texttt{mgf2mm} relative to the Rocket baseline. For the end-to-end workload, \apsmlir achieves a 1.42\x speedup over Rocket. This performance is achieved at a modest implementation cost, with less than 5.3\% area overhead and no frequency degradation.

\compactparagraph{Memory Access Efficiency.} ~\cite{xiao_invited_2025} shows limited gains and even a slowdown (0.21\x). This disparity is attributed to the fully customized dataflow enabled by \apsmlir's enhanced memory bandwidth. As ~\cite{xiao_invited_2025} lacks block-level memory operations, designers intuitively apply scratchpad buffer elision, leading to severe degradation. \apsmlir, however, avoids this pitfall by interface and access pattern analysis. It also correctly assigned these matrix loads to the system bus. These memory access decisions enable \apsmlir to employ more aggressive parallelization strategies, resulting in greater speedups.

\input{table/eval_egraph.tex}

\compactparagraph{Compiler Support.} We intentionally introduced tiling, unrolling, representation transformations like overflow-safe average, and redundant statements in the program code. The compiler successfully matches all variants, demonstrating the effectiveness of our approach. Specifically, the pattern matching process is fully automated, with \apsmlir iteratively applying internal rewrites and external rewrites guided by ISAX loop features. As a result, the overall \egraph size remains manageable, allowing for successful matches within seconds.

\subsection{Point-Cloud Processing}
\label{sec:pcl}
The Iterative Closest Point (ICPAlgo) is a fundamental algorithm for 3D registration, essential for spatial alignment in point-cloud processing (PCP)~\cite{rusu_3d_2011}. We design a set of ISAXs, including \textit{vdist3.vv} (Euclidean distance), \textit{mcov.vs} (covariance matrix), \textit{vfsmax} (maximum comparison), and \textit{vmadot} (matrix-vector multiplication), to accelerate this algorithm. We set the system bus width to 128bit to evaluate whether \apsmlir can effectively utilize the increased bandwidth using interface-aware mechanisms.

\autoref{tab:eval_pcl_deca} details the evaluation results. Compared to the baseline Rocket core, \apsmlir achieves speedups of 1.46\x-9.27\x. For the full pipeline, \apsmlir achieves a 1.96\x speedup. The integration of ISAXs incurs no frequency degradation and a maximum area overhead of 22.9\% for the end-to-end case, which is reasonable given the achieved speedup even for edge devices.

\compactparagraph{Memory Access Efficiency.} The algorithm poses significant memory access challenges due to mixed matrix-vector operations and non-$2^n$ length data accesses. ~\cite{xiao_invited_2025} fails to achieve speedup for the end-to-end case, with bottleneck coming from \textit{vfsmax} and \textit{vmadot} due to suboptimal memory optimization decisions. \apsmlir, however, effectively utilizes the increased bandwidth through the decomposition and scheduling of memory operations, showing the potential for solving real-world design challenges.

\input{fig_tex/eval_pcl_boom}
\compactparagraph{Alternative to General-Purpose Cores.} Upgrading the processor core is a common choice to meet performance demands. However, switching to this much more powerful core will incur a 4.24\x area overhead and 7.3\% frequency drop. We challenge BOOMv3 in handling domain-specific workloads. \autoref{fig:pcl_boom} shows that \apsmlir can be comparable to or even better than BOOM in certain cases with much less area.
Compared to BOOM, where memory traffic is bottlenecked by fixed load-store units (LSUs), \apsmlir can leverage customized dataflow to directly exploit the extended memory bandwidth and drive extensive hardware parallelism.
This proves that \apsmlir can serve as a practical ASIP specialization approach that offers a balanced choice between flexibility and performance.

\compactparagraph{Compiler Robustness. } In the end-to-end case, we intentionally introduced more algebraic transformations to interfere with control flow. As shown in \autoref{tab:egraph-eval}, the compiler continues to match all patterns, demonstrating the effectiveness of our approach. Guided by ISAX loop analysis, loop matching succeeds with limited rewrite passes and keeps the e-node count manageable. In contrast, the attempt to encode entire ISAX patterns as monolithic \egraph rules failed due to excessive complexity and syntactic brittleness. This demonstrates the necessity of our proposed hybrid rewriting strategy, and both internal and external rewrite types are essential and non-interchangeable. Also, this external rewrite mechanism allows the compiler to reuse community-contributed MLIR loop passes, giving us greater confidence in the rewrite correctness.

\subsection{Graphics Rendering}
\label{sec:rendering}

Graphics rendering is another important edge computing application. While these workloads are generally accelerated by existing vector instructions, like the "V" extension of RISC-V, the area induced by such units can be prohibitive for edge applications. We here evaluate whether \apsmlir can achieve better performance-area tradeoffs for specific applications. We design three ISAXs: \textit{vmvar} (1-st and 2-nd moments of vectors), \textit{mphong} (Phong lighting model), and \textit{vrgb2yuv} (color space conversion). We compare them against the Saturn vector unit, configured with VLEN=128.

\input{fig_tex/eval_graphics_rvv}

\compactparagraph{Comparison Result.} Both \apsmlir and Saturn achieve significant speedups over the baseline. \apsmlir achieves speedups of 9.47\x-15.61\x, while Saturn achieves 0.91\x-5.36\x. The poor performance of Saturn on \textit{vmvar} can be attributed to reduction operations, which are inefficient for such instruction sets. For the cycle count alone, Saturn's performance is promising. However, the integration of Saturn results in a 35\% frequency drop, which erodes part of the performance gains and also poses a significant overhead for other parts of the application. \apsmlir also has a much smaller area overhead compared to Saturn(15.6\% and 75\%, respectively). Even if we remove the unused floating-point part from Saturn, \apsmlir can still reduce more than 26\% area. Considering these factors together, \apsmlir achieves a better balance between performance and overhead, making it a better choice for domain specialization.

\input{fig_tex/eval_fpga}
\subsection{CPU LLM Inference}
\label{sec:llm}

We prototype \apsmlir on a physical FPGA platform to validate its real-world performance and hardware efficiency for edge-based LLM inference. We design a set of ISAXs to accelerate attention computation, which is the most computationally intensive part of LLM inference. Our target FPGA is Xilinx XC7Z045, with 350K logic cells, 17.6Mb of BRAM, and 900 DSP slices. \autoref{fig:fpga}(a) shows our test platform. Both cores run at 80MHz and are provided with 1GB DDR3 DRAM, simulating real-world scenarios for edge applications. Under this constraint, we choose the Llama 2 model with 110M parameters, using 8-bit quantization. Synthesis and place-and-route are performed using Vivado 2024.1.

\compactparagraph{Resource Usage and Performance.} \autoref{fig:fpga}(b) demonstrates the resource breakout of the generated SoC. The custom instruction accounts for 15\% LUT, 10\% FF, and 25\% BRAM usage. The relatively high BRAM utilization is primarily dedicated to scratchpad buffers, which mitigates the off-chip memory access bottleneck. This modest resource investment translates into significant performance gains.
Performance results are listed in \autoref{fig:fpga}(c), where \apsmlir achieves 9.30\x and 9.13\x speedup in time-to-first-token (TTFT) and inter-token latency (ITL) compared to the baseline, respectively. This speedup can largely be attributed to the highly efficient memory accesses and the highly parallelized datapath it enables. These results demonstrate the effectiveness of our \apsmlir in real-world scenarios and the practical viability of physical implementation. We believe that with further aggressive quantization and sparsity design, \apsmlir has the potential to achieve even greater speedups.

%% file: table/eval_pcl_deca.tex
\begin{table*}[t]
\caption{Performance and hardware overheads of custom instructions for PQC and PCP workloads.}
\setlength{\tabcolsep}{3pt}
\small
\begin{tabular}{@{}lllccccccccccc@{\hspace{4pt}}}
\toprule
\multirow{2}{*}{Domain}    & \multirow{2}{*}{Case} &  & \multicolumn{3}{c}{Execution Cycle Counts} & \multicolumn{2}{c}{Performance Speedup} &  & \multicolumn{2}{c}{Minimum Clock Period}     &  & \multicolumn{2}{c}{Area Overhead}      \\ \cmidrule(lr){4-6} \cmidrule(lr){7-8} \cmidrule(lr){10-11} \cmidrule(lr){13-14}
 &       &  & Base & ICCAD'25~\cite{xiao_invited_2025} & \textbf{\apsmlir} & ICCAD'25~\cite{xiao_invited_2025} & \textbf{\apsmlir} &  & ICCAD'25~\cite{xiao_invited_2025} & \textbf{\apsmlir} &  & ICCAD'25~\cite{xiao_invited_2025} & \textbf{\apsmlir} \\ \hline
 \multicolumn{1}{l}{\multirow{3}{*}{\shortstack[l]{Post-Quantum\\Cryptography}}} & vdecomp &  & 737    & 189      & 97          & 3.89\x & 7.59\x &  & +0.2\%     & \textcolor{gray}{+0.0\%}         &  & +5.6\%     & +3.2\%         \\
 & mgf2mm  &  & 194    & 928      & 59           & 0.21\x & 3.29\x &  & \textcolor{gray}{+0.0\%}     & \textcolor{gray}{+0.0\%}         &  & +5.3\%     & +1.6\%         \\
 & \textit{End-to-end}       &  & 41314     & 85452       & 28988           & 0.48\x      & 1.42\x      &  & \textcolor{gray}{+0.0\%}     & \textcolor{gray}{+0.0\%}         &  & +11.3\%    & +5.3\%         \\
\hline
\multicolumn{1}{l}{\multirow{5}{*}{\shortstack[l]{Point-Cloud\\Processing}}} & vdist3.vv   &  & 495    & 229      & 137          & 2.16\x & 3.61\x &  & \textcolor{gray}{+0.0\%}     & -0.4\%         &  & +6.9\%     & +8.4\%         \\
\multicolumn{1}{c}{} & mcov.vs &  & 853    & 131      & 92           & 6.51\x & 9.27\x &  & -0.7\%     & \textcolor{gray}{+0.0\%}         &  & +9.6\%     & +9.9\%         \\
\multicolumn{1}{c}{} & vfsmax       &  & 60     & 76       & 41           & 0.79\x & 1.46\x &  & \textcolor{gray}{+0.0\%}     & \textcolor{gray}{+0.0\%}         &  & +9.6\%     & +4.9\%         \\
\multicolumn{1}{c}{} & vmadot       &  & 127    & 202      & 50           & 0.63\x & 2.54\x &  & -0.2\%     & \textcolor{gray}{+0.0\%}         &  & +3.8\%     & +4.2\%         \\
\multicolumn{1}{c}{} & \textit{End-to-end}    &  & 81466  & 99372    & 41596        & 0.82\x & 1.96\x &  & -0.2\%     & \textcolor{gray}{+0.0\%}         &  & +29.6\%    & +22.9\%        \\
\bottomrule
\end{tabular}
\label{tab:eval_pcl_deca}
\end{table*}

%% file: table/eval_egraph.tex
\begin{table}[t]
\centering
\caption{Compilation statistics.}
\renewcommand{\arraystretch}{0.9}
\label{tab:egraph-eval}
\small
\begin{tabular}{@{}l|cc|cc@{}}
\toprule
Case & \makecell{Control-flow\\Difference} & \makecell{Dataflow\\Difference*} & \makecell{Int./Ext.\\Rewrites} & \makecell{Initial/Saturated\\e-nodes} \\
\midrule
vdecomp & Tiling(4) & RF & 8/1 & 298/710 \\
mgf2mm & Unroll(4) & RF, RE & 19/1 & 152/408 \\
\textit{End-to-end} & \makecell{Tiling+Unroll\\[-0.5ex]+Restructure} & RF, RE & 20/2 & 591/1384 \\
\midrule
vdist3.vv & Tiling(8) & AF, RE & 25/1 & 138/700 \\
mcov.vs & Tiling(4) & AF, RF, RE & 24/1 & 187/556 \\
vfsmax & Tiling+Unroll & RF, RE & 15/2 & 263/1212 \\
vmadot & Unroll(2) & RF, RE & 15/1 & 104/359 \\
\textit{End-to-end} & Tiling+Unroll & AF, RF, RE & 30/3 & 696/2842 \\
\bottomrule
\end{tabular}
\leftline{\footnotesize{*AF: Algebraic form, RF: Representation form, RE: Common subexpression split/reuse.}}
\end{table}

%% file: fig_tex/eval_pcl_boom.tex
\begin{figure}[htbp]
  \centering
  \includegraphics[width=\linewidth]{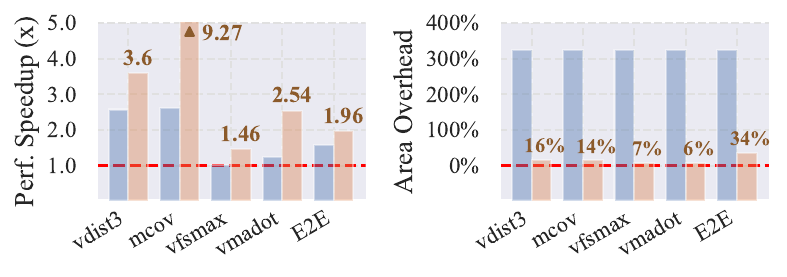}
  \caption{
    Performance and area comparison between \textcolor[rgb]{0.298,0.447,0.690}{BOOMv3} and \textcolor[rgb]{0.867,0.518,0.322}{\apsmlir} on point-cloud processing workloads.
  }
  \label{fig:pcl_boom}
\end{figure}

%% file: fig_tex/eval_graphics_rvv.tex
\begin{figure}[htbp]
  \centering
  \includegraphics[width=\linewidth]{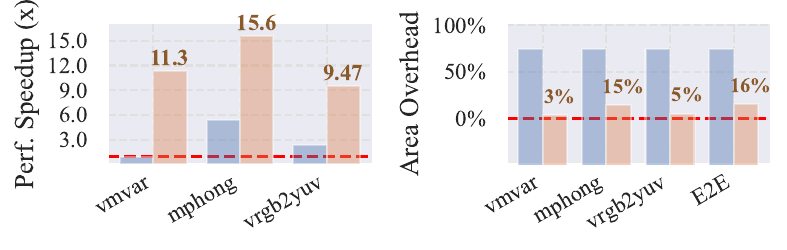}
  \caption{
    Performance and area comparison between \textcolor[rgb]{0.298,0.447,0.690}{Saturn} (RISC-V "V" extension) and \textcolor[rgb]{0.867,0.518,0.322}{\apsmlir} on graphics workloads.
  }
  \label{fig:gfx_rvv}
\end{figure}

%% file: fig_tex/eval_fpga.tex
\begin{figure}[t]
    \centering
    \begin{minipage}{\linewidth}
        \centering
        \includegraphics[width=0.55\linewidth]{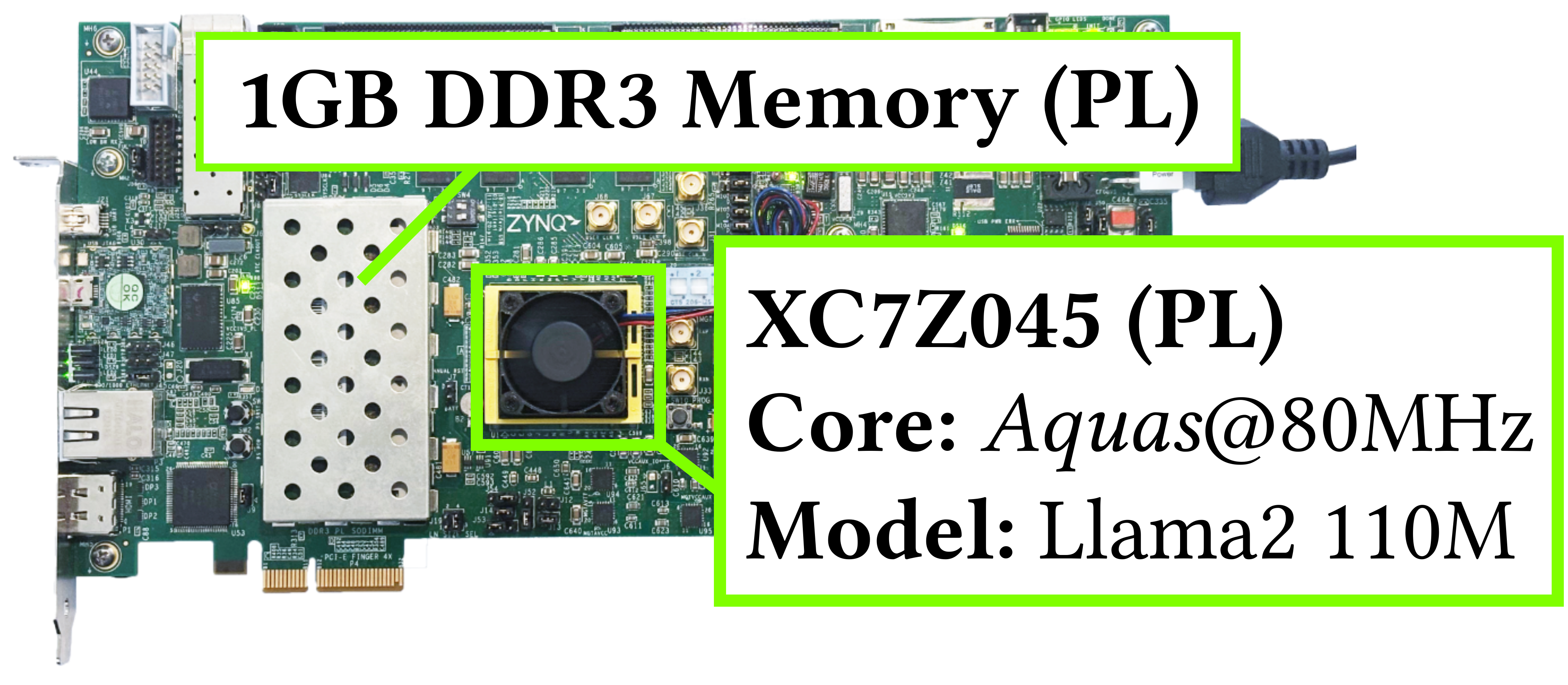}
        \\[-1.2em]
        {\small (a)}
    \end{minipage}
    \begin{minipage}[t][2cm][t]{0.58\linewidth}
        \vspace{0pt}
        \centering
        \includegraphics[width=\linewidth]{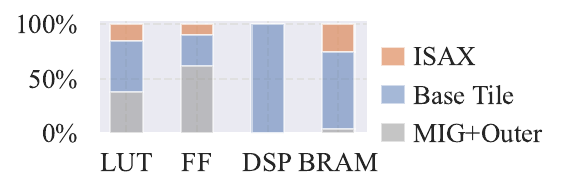}
        \vfill
        {\small (b)}
    \end{minipage}\hfill
    \begin{minipage}[t][2cm][t]{0.39\linewidth}
        \vspace{0pt}
        \centering
        \setlength{\tabcolsep}{3pt}
        \renewcommand{\arraystretch}{0.9}
        \small
        \begin{tabular}{ccc}
            \toprule
            \textbf{Platform} & \textbf{TTFT(s)} & \textbf{ITL(s)} \\
            \midrule
            Base & 32.00 & 32.15 \\
            \textbf{\apsmlir} & 3.44 & 3.52 \\
            \bottomrule
        \end{tabular}
        \vfill
        {\small (c)}
    \end{minipage}

    \caption{FPGA evaluation on LLM inference. (a) the platform setup, (b) the resource breakdown, and (c) latency results.}
    \label{fig:fpga}
\end{figure}

%% file: sec/9_conclusion.tex
\section{Conclusion}

This paper presents \apsmlir, a holistic MLIR-based framework for ASIP hardware-software co-design. Unified in MLIR infrastructure, \apsmlir introduced an interface-aware synthesis flow guided by a model of memory-interface attributes and cache effects for efficient hardware generation, alongside a novel \egraph-based compiler with skeleton-component matching to automate application mapping onto complex ISAXs. Evaluations across four domains show up to 15.61\x kernel speedup and 1.95\x end-to-end acceleration over baseline with no frequency degradation and minor area overheads, showcasing its effectiveness for domain-specific acceleration.